\documentclass{appolb}
\usepackage{graphicx}

\newcommand{\pc}[1]{\ensuremath{\left(#1\right)}}

\usepackage{lscape}
\usepackage{amsmath}
\usepackage{amssymb}
\usepackage{braket}
\usepackage{float}

\begin{document}
\title{Magnetized QCD phase diagram%
\thanks{Presented at {\it Excited QCD}, 7-13 May 2017, Sintra, Portugal}%
}
\author{M\'arcio Ferreira, Pedro Costa, and Constan\c{c}a Provid\^encia, 
\address{CFisUC, Department of Physics, University of Coimbra, P-3004 - 516 Coimbra, Portugal}
}
\maketitle
\begin{abstract}
Using the 2+1 flavor Nambu–-Jona-Lasinio (NJL) model with the Polyakov loop, 
we determine the structure of the QCD phase diagram in an external magnetic field.
Beyond the usual NJL model with constant couplings, we also consider 
a variant with a magnetic field dependent scalar coupling, which
reproduces the Inverse Magnetic Catalysis (IMC) at zero chemical potential.
We conclude that the IMC affects the location of the Critical-End-Point, and found
 indications that, for high enough magnetic fields, the chiral phase transition at zero chemical potential
 might change from an analytic to a 
first-order phase transition.
\end{abstract}
  
\noindent\textit{\textbf{Introduction}:} The properties of hadronic matter in a magnetized environment
is attracting the attention of the physics community.
The effect of an external magnetic field on the chiral and deconfinement
transitions is an active field 
of research with possible relevance in multiple physical systems. 
From heavy-ion collisions at very high energies, 
to the early stages of the Universe 
and astrophysical objects like magnetized neutron stars, 
the magnetic field may play an important role.  

The catalyzing effect of an external magnetic field on dynamical chiral 
symmetry breaking, known as Magnetic Catalysis (MC) effect, is well understood \cite{Miransky:2015ava}. 
However, Lattice QCD (LQCD) studies show  an additional effect \cite{baliJHEP2012,bali2012PRD,endrodi2013}, 
the Inverse Magnetic Catalysis (IMC): instead of catalyzing, the magnetic field weakens the dynamical chiral symmetry 
breaking in the crossover transition region. 
The chiral pseudo-critical transition temperature turns out to be a decreasing function of the
magnetic field strength. 

Different theoretical approaches have been applied in studying the magnetized QCD phase diagram,
and specifically the IMC effect. 
Several low-energy effective models, including the Nambu--Jona-Lasinio (NJL)-type models, 
have been used to investigate the impact of external magnetic fields on quark matter
(for a recent review see \cite{Andersen:2014xxa}).\\

\vspace{-0.2cm}
\noindent\textit{\textbf{Model}:} We perform our calculations in the framework of the Polyakov--Nambu--Jona-Lasinio 
(PNJL) model. The Lagrangian in the presence of an external magnetic field is 
given by
\begin{align*}
{\cal L} &= {\bar{q}} \left[i\gamma_\mu D^{\mu}-{\hat m}_f \right ] q + 
	G_s \sum_{a=0}^8 \left [({\bar q} \lambda_ a q)^2 + ({\bar q} i\gamma_5 \lambda_a q)^2 \right ]- \frac{1}{4}F_{\mu \nu}F^{\mu \nu}\\
	&-K\left\{{\rm det} \left [{\bar q}(1+\gamma_5) q \right] + 
	{\rm det}\left [{\bar q}(1-\gamma_5)q\right]\right\} + 
	\mathcal{U}\left(\Phi,\bar\Phi;T\right),
	\label{Pnjl}
\end{align*}
where $q = (u,d,s)^T$ represents a quark field with three flavors, 
${\hat m}_f= {\rm diag}_f (m_u,m_d,m_s)$ is the corresponding (current) mass 
matrix, and $F_{\mu \nu }=\partial_{\mu }A^{EM}_{\nu }-\partial _{\nu }A^{EM}_{\mu }$ is 
the (electro)magnetic tensor.
The covariant derivative $D^{\mu}=\partial^\mu - i q_f A_{EM}^{\mu}-i A^\mu$
couples the quarks to both the magnetic field $B$, {\it via} $A_{EM}^{\mu}$, and 
to the effective gluon field, {\it via} $A^\mu(x) = g{\cal A}^\mu_a(x)\frac{\lambda_a}{2}$, where
${\cal A}^\mu_a$ is the SU$_c(3)$ gauge field.
The $q_f$ represents the quark electric charge ($q_d = q_s = -q_u/2 = -e/3$).
We consider a  static and constant magnetic field in the $z$ direction, 
$A^{EM}_\mu=\delta_{\mu 2} x_1 B$. We employ the logarithmic effective potential 
$\mathcal{U}\left(\Phi,\bar\Phi;T\right)$ \cite{Roessner:2006xn}, fitted
to reproduce lattice calculations. 

We use a sharp cutoff ($\Lambda$) in three-momentum 
space as a model regularization procedure.  
The parameters of the model are \cite{Rehberg:1995kh}: $\Lambda = 602.3$ MeV, 
$m_u=m_d=5.5$ MeV, $m_s=140.7$ MeV, $G_s^0 \Lambda^2= 1.835 $ and 
$K \Lambda^5=12.36$.

We analyze two model variants with distinct
scalar interaction coupling: a constant coupling 
$G_s=G_s^0$ and a magnetic field dependent coupling $G_s=G_s(eB)$ \cite{Ferreira:2014kpa}.
In the latter, the magnetic field dependence
is determined phenomenologically, by reproducing the decrease ratio of the
chiral pseudo-critical temperature obtained in LQCD calculations \cite{baliJHEP2012}.
Its functional dependence is
$G_s(\zeta)=G_s^0\pc{\frac{1+a\,\zeta^2+b\,\zeta^3}{1+c\,\zeta^2+d\,\zeta^4}}$,
where $\zeta=eB/\Lambda_{QCD}^2$ (with $\Lambda_{QCD}=300$ MeV).
The parameters are $a = 0.0108805$, $b=-1.0133\times10^{-4}$, $c= 0.02228$, and $d=1.84558\times10^{-4}$ \cite{Ferreira:2014kpa}.\\

\begin{figure}[!htbp]
  \centering
  	\includegraphics[width=0.43\linewidth]{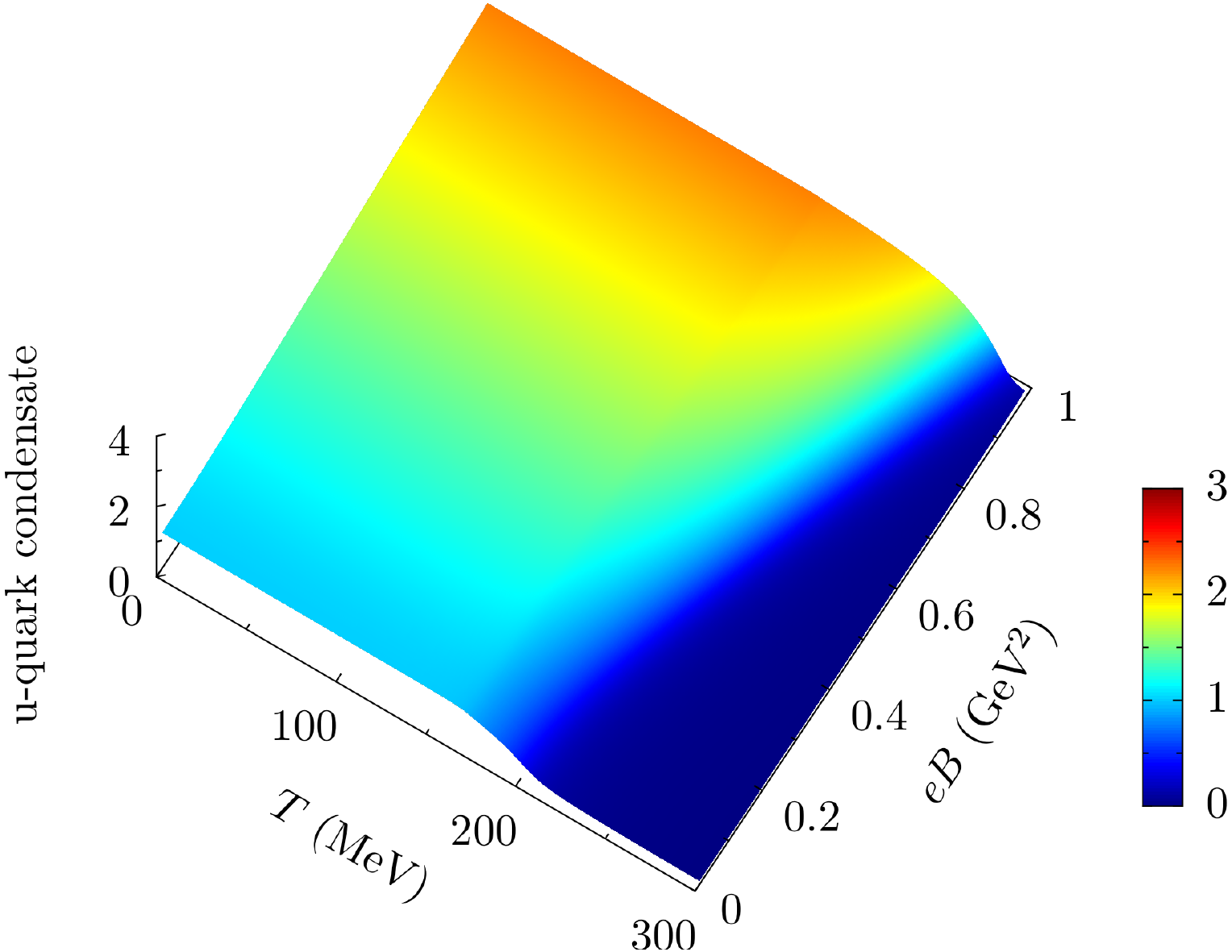}
  	\hspace{10pt}
  	\includegraphics[width=0.43\linewidth]{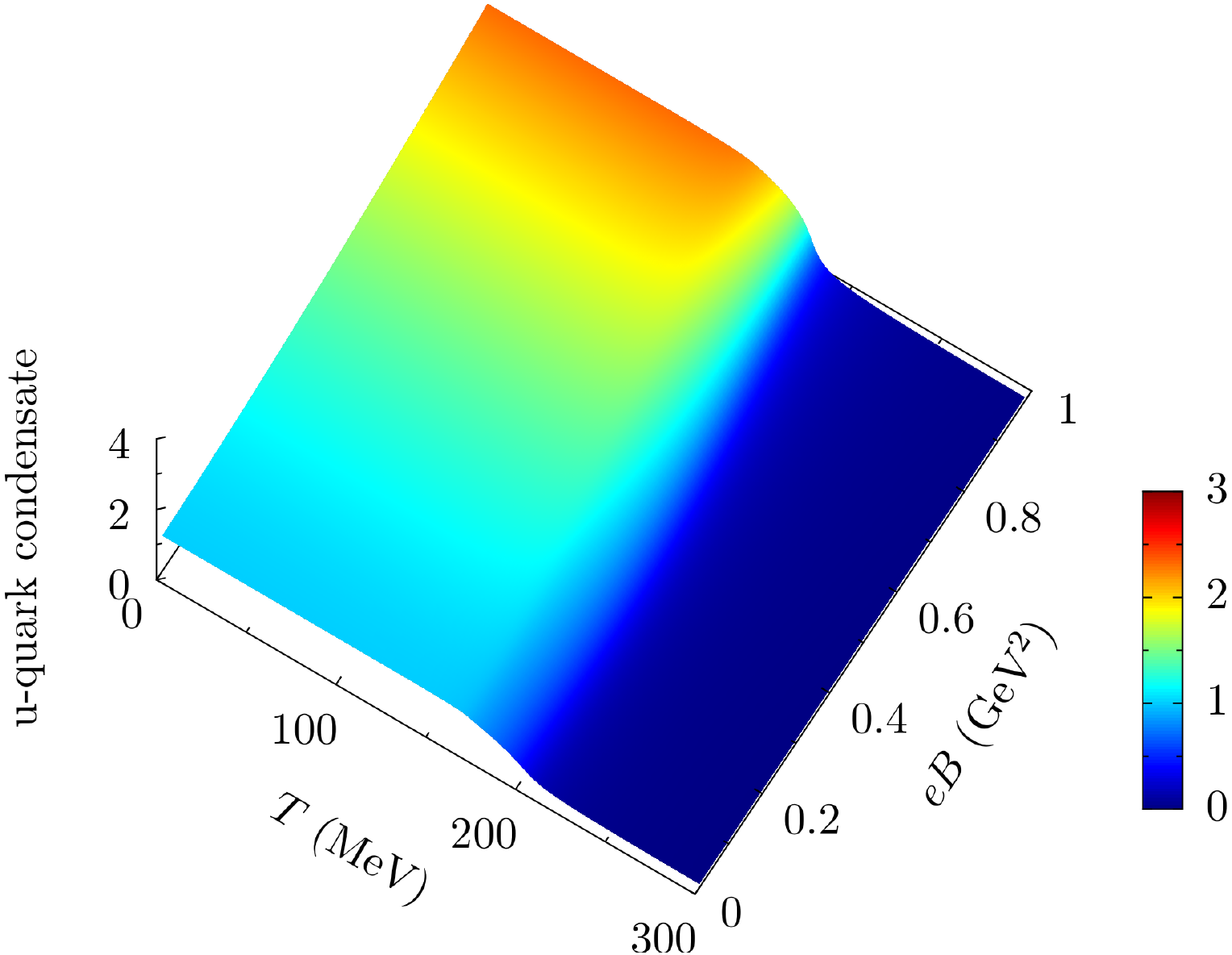}\\
  	\includegraphics[width=0.43\linewidth,angle=0.0]{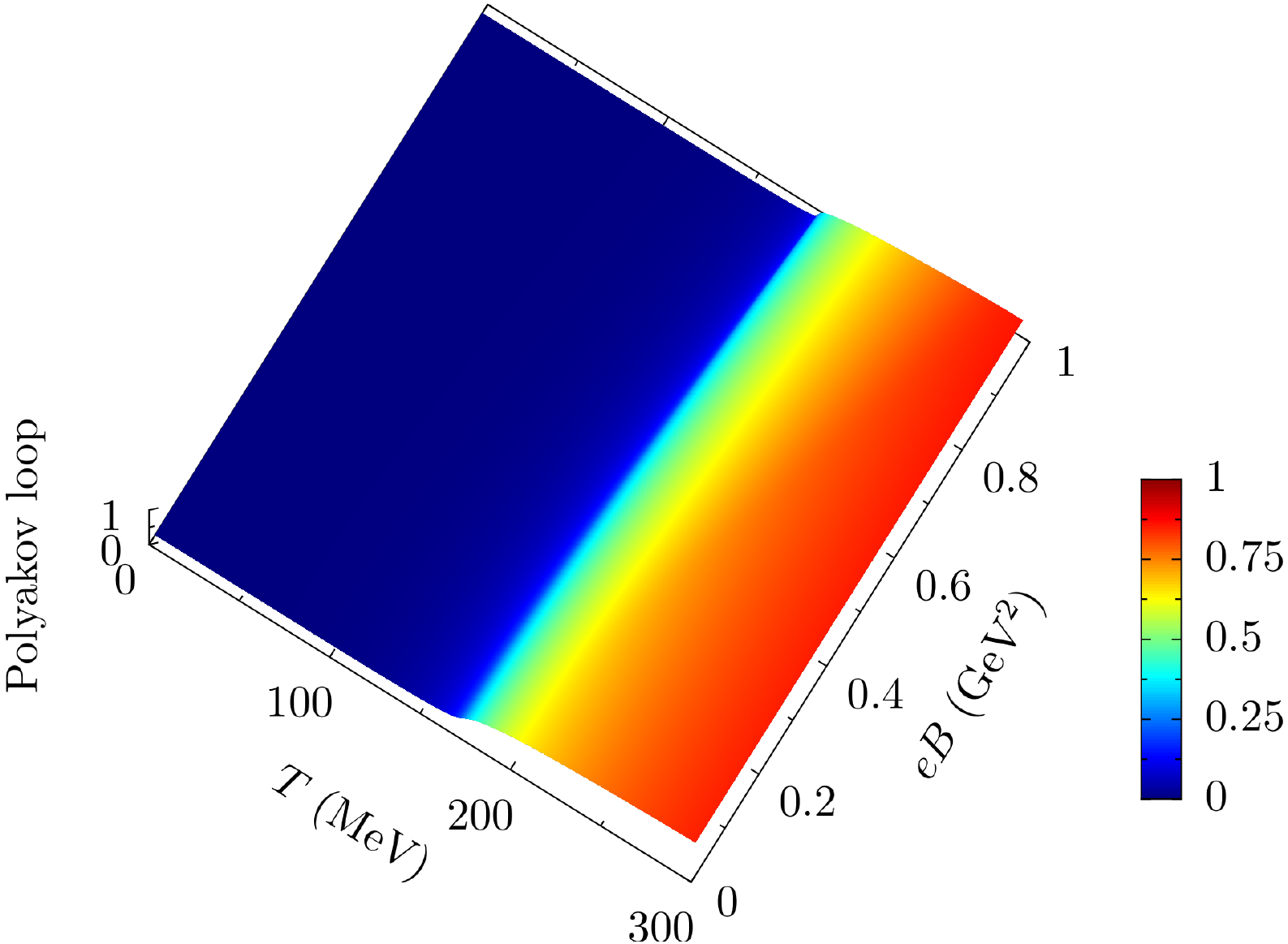}
  	\hspace{10pt}
  	\includegraphics[width=0.43\linewidth,angle=0.0]{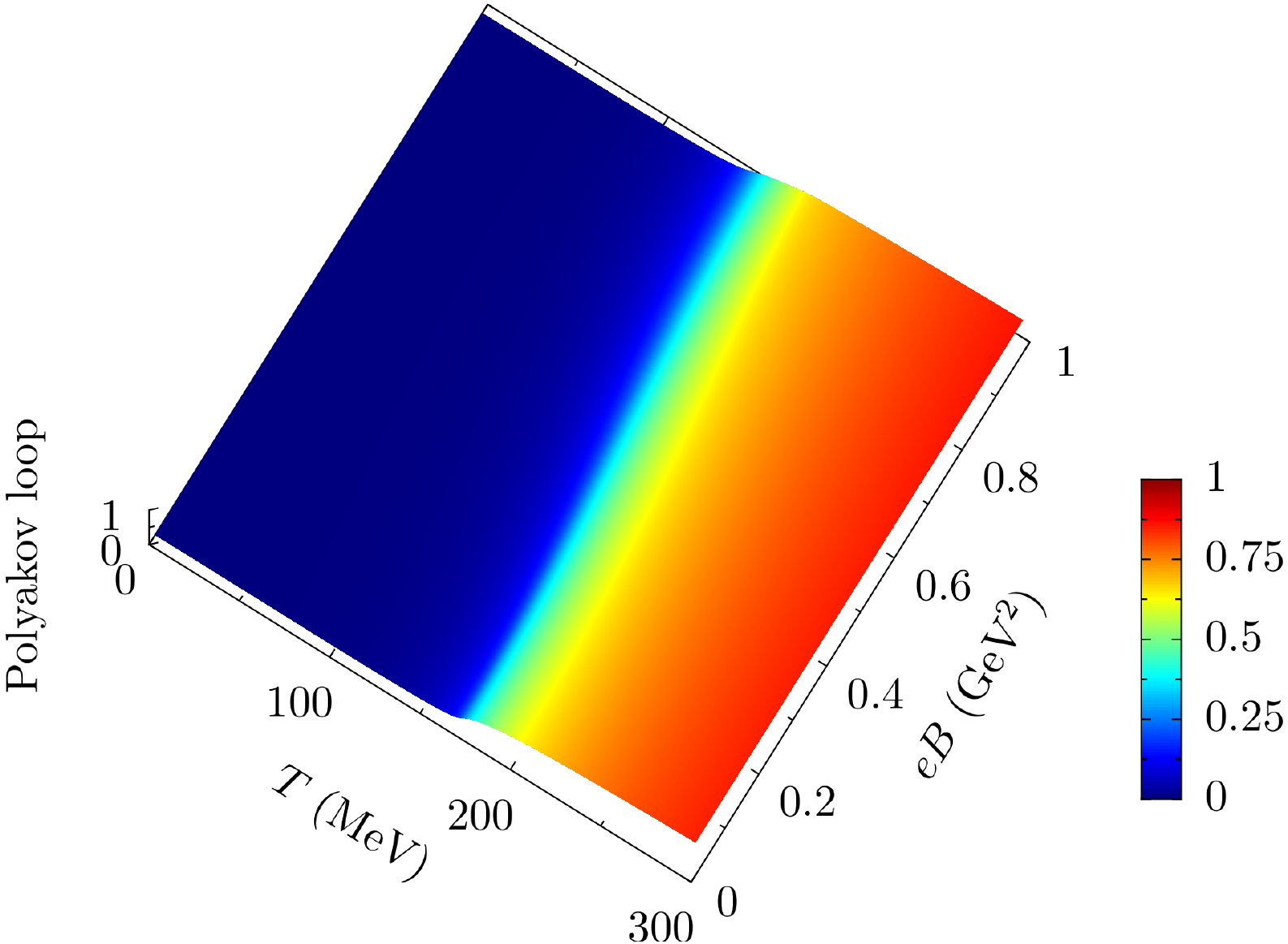}
\caption{Vacuum normalized u-quark condensate (top) and
	Polyakov loop value (bottom)
	 for $G_s^0$ (left) and $G_s(eB)$ (right).}
\label{fig:1}
\end{figure}

\vspace{-0.2cm}
\noindent\textit{\textbf{Results (zero chemical potential)}:} Let us first compare both models at zero chemical potential. 
The up-quark condensate (all quarks show similar results), 
normalized by its vacuum value, and the 
Polyakov loop value are in Fig. \ref{fig:1}. 
The presence of the IMC effect in the $G_s(eB)$ model its clear 
in Fig. \ref{fig:1} (right top panel), by the suppression effect of the magnetic field on 
the quark condensate around the transition temperature region. 
Furthermore, the $G_s(eB)$ model still leads to Magnetic Catalysis at low and high temperatures: 
the magnetic field enhances the quark condensate away from the transition temperature
region, i.e., at low and high temperatures. 
The chiral pseudo-critical transition temperature, defined as the inflection point of the quark condensate,
decreases for $G_s(eB)$ and increases for $G_s^0$.  
The $G_s(eB)$ makes possible not only the decreasing transition temperature, but also preserves the analytic nature 
of the chiral transition, in accordance with LQCD results.
The $G_s(eB)$ dependence also affects the Polyakov loop value (bottom panel).
A decreasing pseudo-critical temperature for the deconfinement transition with increasing magnetic field
is obtained
for $G_s(eB)$, contrasting with the increasing pseudo-critical temperature
for $G_s^0$. The $G_s(eB)$ dependence induces a reduction of the Polyakov loop
value in the transition temperature region (also seen in LQCD results \cite{endrodi2013}). \\

\begin{figure}[!htbp]
	\centering
	\includegraphics[width=0.8\linewidth,angle=0.0]{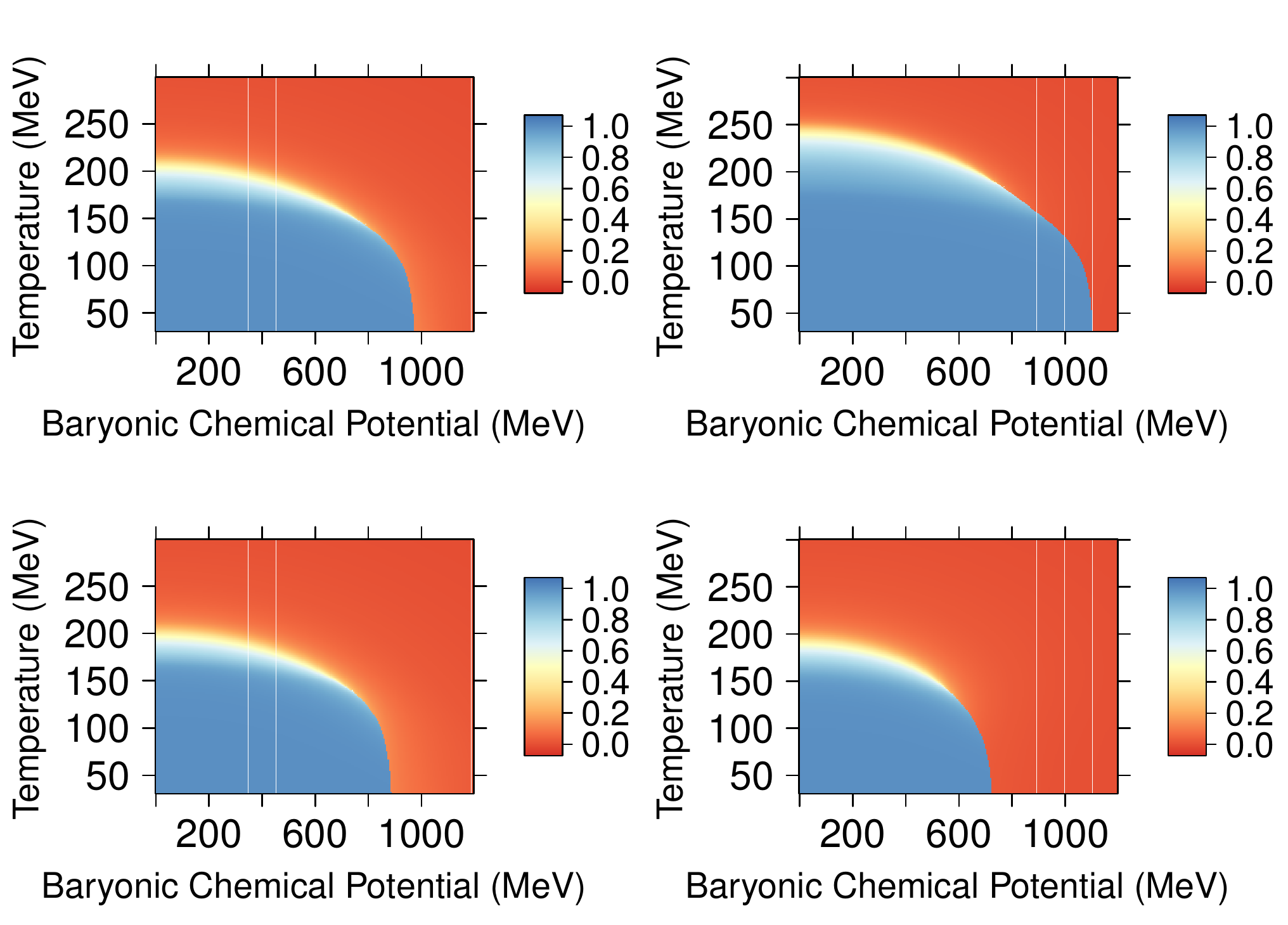}
	\caption{Up-quark condensate (normalized by its vacuum value) with $G_s^0$ (top) and $G_s(eB)$ (bottom)
		for $eB=0.2$ GeV$^2$ (left) and $eB=0.6$ GeV$^2$ (right). 
		The color scale represents the magnitude of the vacuum normalized condensate.}
	\label{fig:2}
\end{figure}

\begin{figure}[!htbp]
	\centering
	\includegraphics[width=0.8\linewidth,angle=0.0]{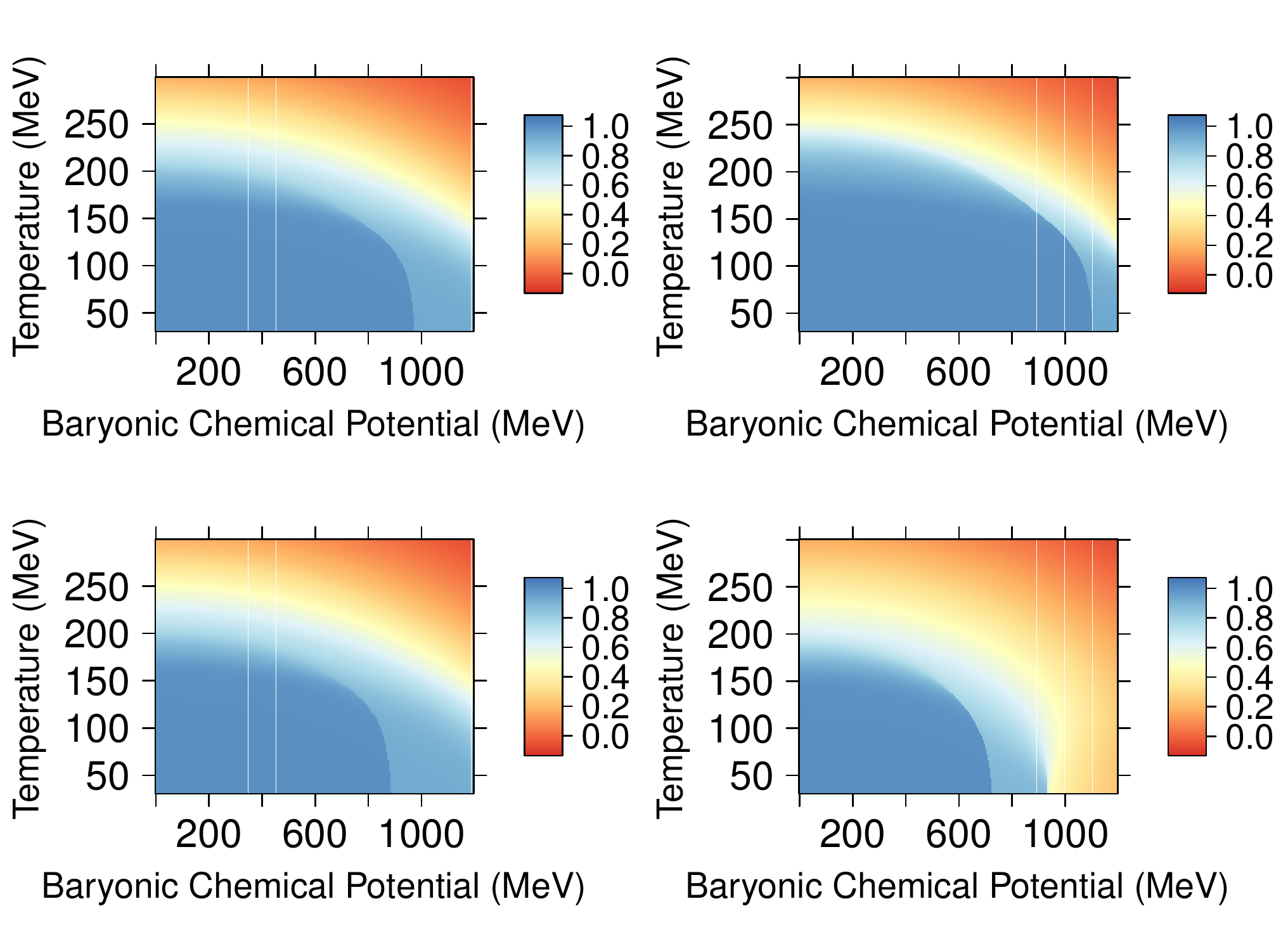}
	\caption{Strange-quark condensate (normalized by its vacuum value) with $G_s^0$ (top) and $G_s(eB)$ (bottom)
		for $eB=0.2$ GeV$^2$ (left) and $eB=0.6$ GeV$^2$ (right).
	The color scale represents the magnitude of the vacuum normalized condensate.}
	\label{fig:3}
\end{figure}

\begin{figure}[!htbp]
	\centering
	\includegraphics[width=0.8\linewidth,angle=0.0]{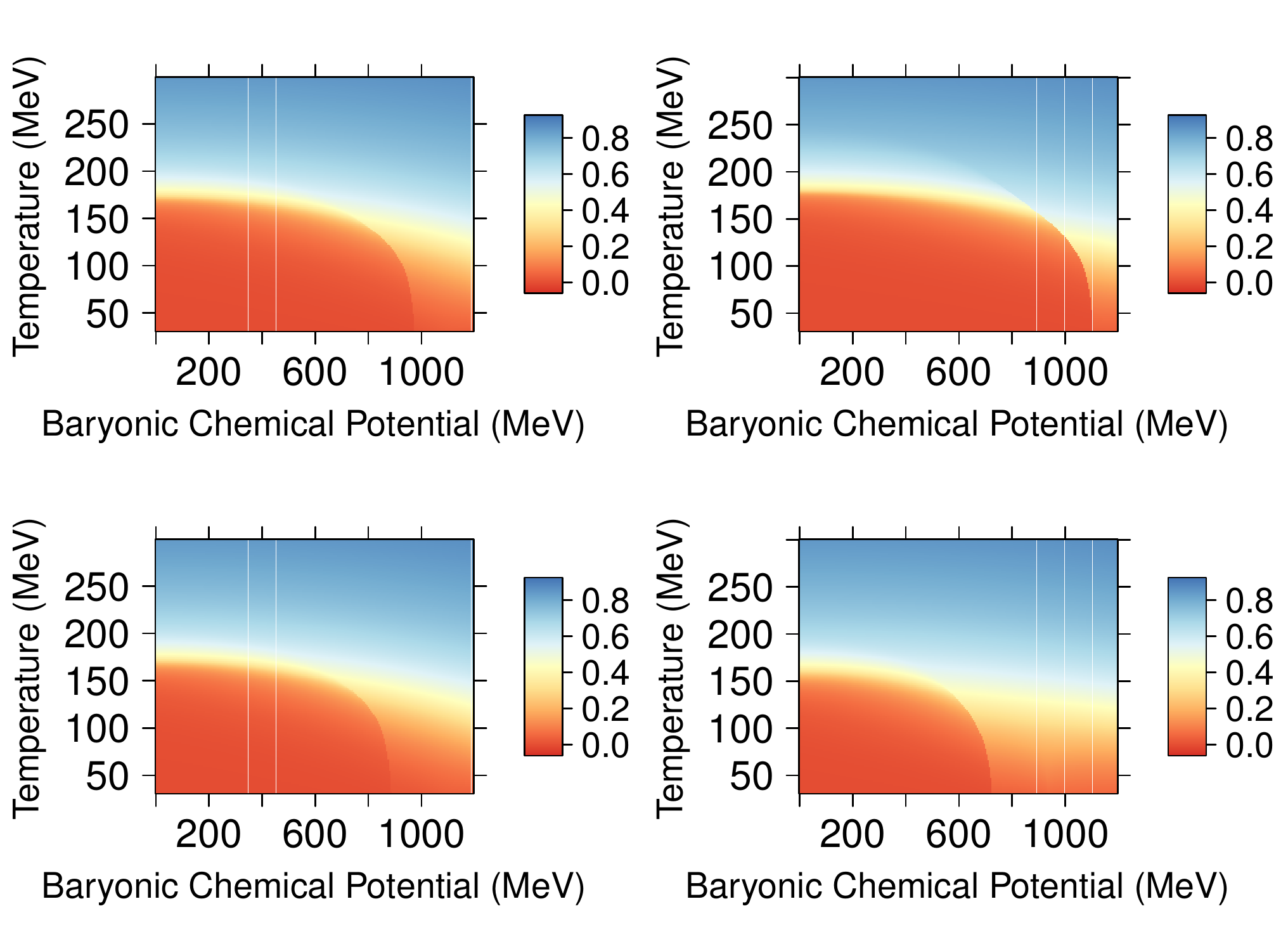}
	\caption{Polyakov loop value $\Phi$ with $G_s^0$ (top) and $G_s(eB)$ (bottom)
		for $eB=0.2$ GeV$^2$ (left) and $eB=0.6$ GeV$^2$ (right).
	The color scale represents the Polyakov loop magnitude.}
	\label{fig:4}
\end{figure}

\vspace{-0.2cm}
\noindent\textit{\textbf{Results (finite chemical potential)}:} Now, by introducing a finite chemical potential, we analyze the impact of the
$G_s(eB)$ on the entire phase diagram. The results are displayed
in Figs. \ref{fig:2}, \ref{fig:3}, and \ref{fig:4}, where 
the respective quantities are presented
for two magnetic field intensities ($0.2$ GeV$^2$ and $0.6$ GeV$^2$) 
within both models.
From Fig. \ref{fig:2}, we see that the (partial) chiral restoration is accomplished  
via an analytic transition (crossover) at low chemical potentials, and
through a first-order phase transition at higher chemical potentials.
The region on which the chiral phase is broken (blue region) shrinks as the
magnetic field increases for the $G_s(eB)$ model, and the opposite occurs for $G_s^0$.  
Similar plots are shown in Fig. \ref{fig:3}, but now for the strange quark.
The general pattern shows a smoothly decrease of the strange quark condensate 
over the whole phase diagram, though some discontinuities appear, which are induced by 
the first-order phase transition of the light quarks. An interesting result 
is seen for the $G_s(eB)$ model at $eB=0.6$ GeV$^2$ (bottom right panel of Fig.~\ref{fig:3}): 
a first-order phase transition shows up for the strange quark at
low temperatures which ends up in a Critical-End-Point (CEP) at a temperature around $50$ MeV.
Finally, we represent the Polyakov value in Fig. \ref{fig:4}. 
The general pattern is maintained within both models.
We see that the transition from confined quark matter ($\Phi\approx0$)
to deconfinement quark matter ($\Phi\approx1$) is accomplished
via an analytic transition, reflected in the continuous increase 
of the Polyakov loop value (there is a discontinuity induced 
by the chiral first-order phase transition, on which the variation 
of the Polyakov loop value is small). Because the chiral broken phase
region gets smaller with increasing magnetic field, the
region on which the chiral phase is (approximately) restored but still confined
(at low temperatures and high chemical potentials)
enlarges with increasing magnetic field strength. 
The opposite occurs for the model with constant coupling.

\begin{figure}[!htbp]
  \centering
  	\includegraphics[width=0.5\linewidth,angle=0.0]{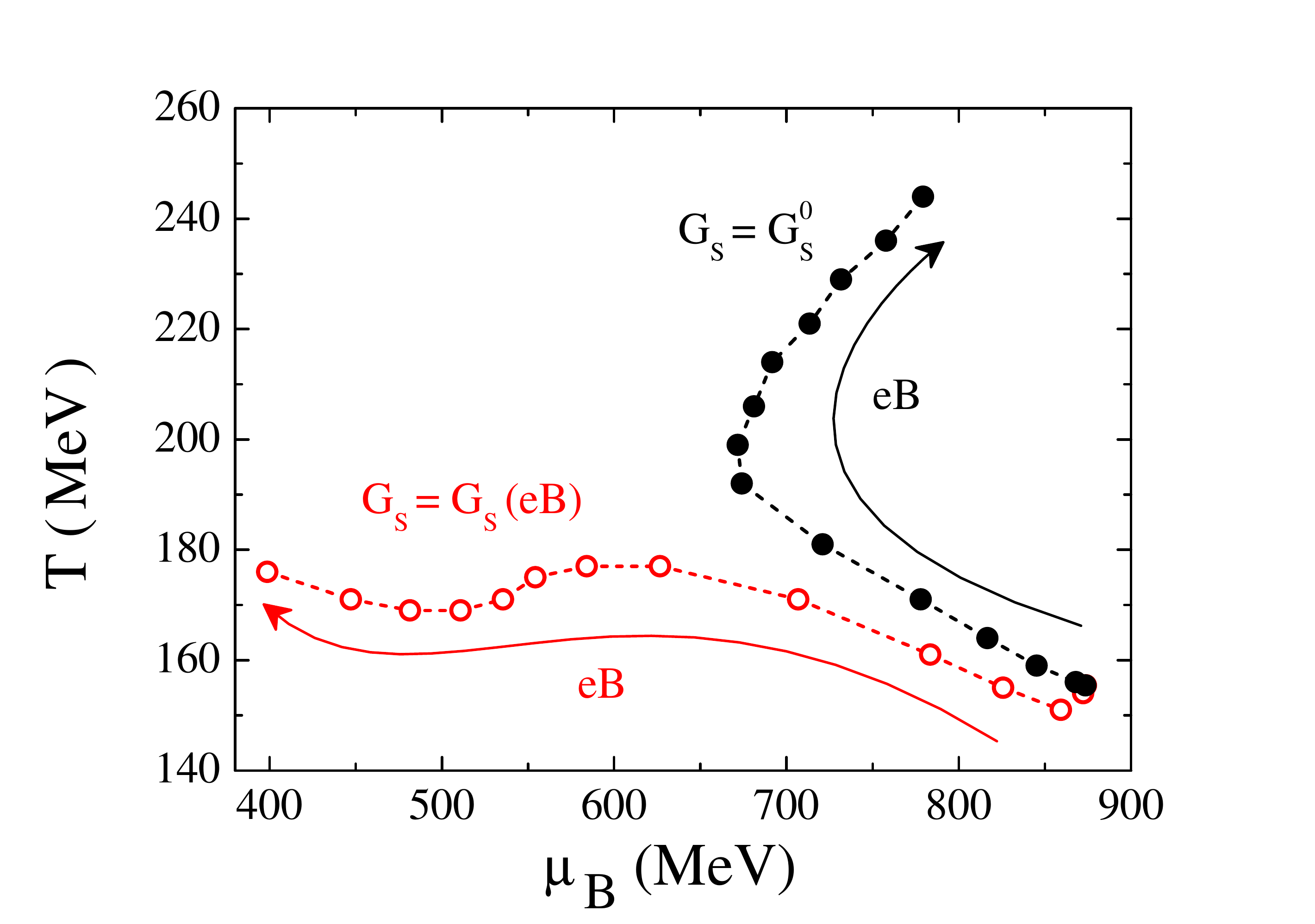}
  	\caption{The CEP position with increasing $B$ field 
  	for $G_s^0$ (black) and $G_s(eB)$ (red).}
\label{fig:5}
\end{figure}

As a final step, we focus on the CEP's location of the chiral transition as a function
of the magnetic field \cite{Costa:2013zca,Costa:2015bza}. The result is shown in Fig.~\ref{fig:5}.
An important result shows up that clearly differentiates both models. 
Despite the agreement at low magnetic field strengths ($eB<0.1$) between both models
on how the CEP reacts to the $B$ presence, for higher magnetic fields the
CEP moves towards lower chemical potentials for $G_s(eB)$, while it moves 
for higher chemical potentials for $G_s^0$. This might indicate
that for high enough magnetic fields, the chiral phase transition 
might change from an analytic to a first-order phase transition at 
zero chemical potential
 (there are some indications for this scenario \cite{Endrodi:2015oba}). 
\vspace{-0.3cm}
\vspace{-0.3cm}

{\bf \it Acknowledgments}:
This work was partly supported by Project PEst-OE/FIS/UI0405/2014 
developed under the initiative QREN financed by the UE/FEDER through the
program COMPETE $-$ ``Programa Operacional Factores de
Competitividade'', and by Grants No. SFRH/BD/\-51717/\-2011 and No.  
SFRH/BPD/1022 73/2014 from F.C.T., Portugal.

\end{document}